\documentclass[12pt]{article}
\usepackage{latexsym}
\usepackage{bm}
\newcommand{\beq}{\begin{equation}}
\newcommand{\eeq}{\end{equation}}
\newcommand{\bc}{\begin{center}}
\newcommand{\ec}{\end{center}}
\newcommand{\eeqa}{\end{eqnarray}}
\newcommand{\beqa}{\begin{eqnarray}}
\newcommand{\no}{\noindent}
\newcommand{\pa}{\partial}

\newcommand{\ra}{\rightarrow}
\newcommand{\na}{\nabla}
\newcommand{\foot}{\footnotemark\ }
\newcommand{\al}{\alpha}
\newcommand{\be}{\beta}
\newcommand{\ga}{\gamma}
\newcommand{\Ga}{\Gamma}
\newcommand{\de}{\delta}

\newcommand{\ep}{\epsilon}
\newcommand{\ze}{\zeta}
\newcommand{\et}{\eta}

\newcommand{\si}{\sigma}

\newcommand{\ta}{\tau}

\newcommand{\ph}{\phi}

\newcommand{\ps}{\psi}

\newcommand{\ed}{\end{document} }

\begin{document}

\title{Strings in gravity with torsion}

\author{Richard T. Hammond\thanks{e-mail rhammond@plains.nodak.edu}\\
North Dakota State University\\ Physics Department\\Fargo, North
Dakota 58105 U.S.A.}
\date{\today}
\maketitle
\begin{abstract}
A theory of gravitation in 4D is presented with strings
used in the material action in $U_4$ spacetime. It is shown that
the string naturally gives rise to torsion. It is also shown that the equation
of motion a string follows from the Bianchi identity, gives the identical
result as the Noether conservation laws, and follows a geodesic
only in the lowest order approximation. In addition, the
conservation laws show that strings naturally have spin, which arises
not from their motion but from their one dimensional structure.

\end{abstract}


\section{Introduction}
Two branches in general relativity have met with varying
degrees of success over the years. One is the use of strings
to model, or reflect, the structure of matter. This topic,
while still relatively young, has provided substantial hope
into a successful program of quantum gravity,\foot and
generates great interest at the classical level as well.\foot
%ignites many intriguing putative solutions to
The other branch reaches further back and modifies Einstein's
original theory by taking the affine connection to be
asymmetric. The antisymmetric part of the affine
connection --- the torsion, has a twisted history,
but gained renewed popularity in the 1970s when it was
used in developing a local Poincar\`e gauge theory
of gravity.\foot The interest in torsion gained a wider audience
when it was shown that it could could act as the antisymmetric
field that is required in string theory,\foot
and gained further
notice when it appeared  in supergravity theories.\foot
The two once disparate branches have  become intertwined,
and the purpose here is to tap into the synergism
generated by this growth.

One of the underlying bonds that seems to unite string theory
and general relativity with torsion is the assumption
that the torsion may be derived from an antisymmetric
potential $\ps_{\mu\nu}$ according to
\beq\label{tordef}
S_{\mu\nu\si}=\ps_{[\mu\nu,\si]}
.\eeq
When this assumption was used to develop a theory of gravitation
with torsion, the physical interpretation that fell upon torsion
was that it was created from intrinsic spin.\foot The interpretation
resulted from an analysis of the conservation laws  of the
equations of motion. Some of the salient features that were
discussed and shown in \cite{sg} are the following. It was shown
that it was necessary to introduce an intrinsic vector $\xi^\mu$
to represent the source for torsion. This may be viewed as a generalization
of other intrinsic quantities used in describing sources. For
example, to describe a point particle we may consider its mass
$m$ as an intrinsic quantity, or its charge $q$ as an
intrinsic quantity. The source tensors are built up from these
quantities. In order to couple a source to an antisymmetric
field, as is the case for torsion, it is necessary
to go beyond scalar intrinsic quantities and adopt the intrinsic
vector approach. This gave rise to several aspects of the theory.
One is that intrinsic spin is derived from this intrinsic
vector. Thus, intrinsic spin arises, not from motion or rotation,
but from structure and persists in the limit of a static configuration.
Another result was that the particle
had to have structure, and labeling each point on the structure
by a vector $\xi^\mu_n$, it was shown that the conservation laws
imply $\sum_n\xi^\mu_n=0$. The conservation laws also
predicted a definite spin interaction term, that in principle,
could be observed. This interaction term was confirmed when
the Dirac Lagrangian was used in place of the intrinsic
vector material action. In the low energy limit, the Dirac equation
yielded that same interaction as that predicted by the intrinsic
vector approach,\foot and this was used to place an upper
bound on the coupling constant.\foot
In addition, it was shown that a scalar
arises naturally, \foot and it was also shown that this scalar
field might be interpreted as the scalar field of string theory.\foot

Thus, the theory of gravitation with torsion with (\ref{tordef})
adopts many characteristics of string theory --- from the mirrored
Lagrangian of the low energy string theory limit, to a source
which must be at least one dimensional. It is natural then to
consider replacing the intrinsic vector material Lagrangian by
that of a string. In this case, many plausible results follow.
First, the intrinsic vector will be seen to correspond to the
tangent vector of the string. Also the condition given above that
the intrinsic vectors sum to zero, is replaced by the much more
natural result $\oint d\ze=0$, which is true for closed strings.
Moreover, there is a natural coupling between the torsion
potential and the area swept out by the string. It will be shown
that many of the results of the intrinsic vector approach will be
reproduced by use of the string source, and in fact, in many ways,
do  so in a much more natural fashion. It will also be shown that
strings have intrinsic spin, which follows from their structure
(not motion), and the equations of motion will follow from the
conservation laws. In what follows only closed strings will be
considered, the discussion of open strings will be considered
separately. Finally, I would like to give one more preview of
things to come. It is customary to obtain the equation of motion
of a string by adopting the geodesic postulate, i.e., setting the
variations of the action with respect to the coordinate equal to
zero, or in a more rigorous approach, using invariance of the
material action.\foot However, in general relativity the equations
of motion of an object follow from the field equations, and
variations with respect to the coordinate will not be considered.
Thus, certain conditions or properties of the string well known
from string theory are not necessarily appropriate to this case.

\section{Field equations and the conservation laws}
\subsection{Field equations}
The material in this section is general in the sense that
no explicit form of the energy momentum tensor is given.
Details on this material may be found in \cite{sg}.
The field equations are given by the action principle
\beq
\de (I_g +I_m)=0
\eeq
where

\beq\label{R}
I_g=\int\sqrt{-g}{R\over2k}d^4x
,\eeq
 $I_m$ is the material action, and $ k=8\pi$.
The curvature scalar $R$ is that of $U_4$ spacetime,
and therefore contains torsion.  The
unknown quantities are the 10 components of the symmetric metric
tensor and the 6 components of the antisymmetric torsion
potential. The equations are obtained by considering independent
variations of these potentials. (One may note that this action
principle yields second order differential equations in the torsion
potential, whereas
an action given by (\ref{R}), but with variations taken with respect
to the torsion itself, yields non-propagating torsion.)
However, an equivalent procedure
is, defining $\ph_{\mu\nu}= g_{\mu\nu}+\ps_{\mu\nu}$, to consider
variations with respect to $\ph_{\mu\nu}$. Now, taking variations
with respect to $\ph_{\mu\nu}$ gives  a set of 16 equations. The
symmetric part is equivalent to those obtained by considering
variations with respect to the metric tensor, and will be called
the gravitational field equations. The antisymmetric part is
equivalent to the equations obtained by performing variations with
respect to the torsion potential and will be called the torsional
field equations. In this case, a non-symmetric energy momentum
tensor was defined (see \cite{sg}) according to

\beq\label{emt}
\de I_m={1\over2}\int d^4x\sqrt{-g}T^{\mu\nu}\de \ph_{\mu\nu}
.\eeq
The symmetric part of the energy momentum tensor in
(\ref{emt}) is the source in the gravitational field equations,
and the antisymmetric part is the source in the torsional field
equations. With these remarks, the field equations are given by

\beq\label{fe}
G^{\mu \nu}  - 3S^{\mu \nu \sigma}_{\ \ \ \ ; \sigma}
-2S^{\mu}_{\ \alpha \beta}S^{\nu \alpha \beta} = kT^{\mu \nu}
\eeq
where
\beq
G^{\mu\nu}=R^{\mu\nu}-{1\over2}g^{\mu\nu}R
\eeq
and $R^{\mu\nu}$ is the (asymmetric) Ricci tensor in $U_4$ spacetime.
In line with the above remarks, the torsional field equations
are given by

\beq\label{tfe}
S^{\mu \nu \sigma}_{\ \ \ \ ;\sigma} =-kj^{\mu \nu}
\eeq
where $j^{\mu\nu}\equiv (1/2)T^{[\mu\nu]}$.
In the above, and below, it is convenient to use two
different kinds of covariant differentiation. First, the
fundamental definition of a covariant derivative is given
by, for any vector $A^\mu$,

\beq
\na_\si A^\mu=A^\mu_{\ ,\si}+ \Ga_{\si \nu}^{\ \ \ \mu}A^\nu
\eeq
which  contains the full (asymmetric) affine connection
$\Ga_{\si \nu}^{\ \ \ \mu}$.
However, sometimes the antisymmetric part drops out, and it is useful
to define the ``Christofell'' covariant derivative by

\beq
A^\mu_{\ ;\si}=A^\mu_{\ ,\si}+ \{_{\si \nu}^{\; \mu}\}A^\nu
\eeq
 where $\{_{\si \nu}^{\; \mu}\}$ is the (symmetric)
Christofell symbol. The relation between the affine connection
and the Christoffel symbol is obtained by the requirement
$\na_\si g_{\mu\nu}=0$, which yields
\beq
\Ga_{\mu \nu}^{\ \ \ \si}=\{_{\mu \nu}^{\; \si}\}
+S_{\mu\nu}^{\ \ \ \si}+S^\si_{\ \mu\nu}+S^\si_{\ \nu\mu}
\eeq
which becomes, with (\ref{tordef}),
\beq
\Ga_{\mu \nu}^{\ \ \ \si}=\{_{\mu \nu}^{\; \si}\}
+S_{\mu\nu}^{\ \ \ \si}
.\eeq

\subsection{Conservation laws}
Ultimately, the  correctness of any theory can only
be ascertained by experiment. In a theory
of gravitation this means that one must derive
the equations of motion, which predict acceleration,
interaction energy or something that is measurable.
Not only do the equations of motion provide the
means to evaluate the theory, they are also
instrumental in developing the physical interpretation of the theory.
A powerful aspect of general relativity is that the
equations of motion follow from the field equations,
and therefore already establish the machinery for the analysis
of the predictions of the theory. In fact, this comes about in
two ways, each of which yields the same result. One is that
the Bianchi identities must be obeyed, which are,
in $U_4$ spacetime,

\begin{equation}\label{bi}
\nabla_{\nu}G^{\mu \nu} = 2S^{\mu \alpha \beta}R_{\beta \alpha}
-S_{\alpha \beta \gamma}R^{\mu \gamma \beta \alpha}
.\end{equation}
To use the Bianchi identities, one operates with
$\na_\nu$ on (\ref{fe}) and uses (\ref{bi}). This establishes
a differential relation on the source. The other way to proceed
is to capitalize on the requirement that the material
action is a scalar, and that under the manifold mapping
$x^\mu\ra x^\mu+\ep^\mu$,
\beq
\int d^4x\sqrt{-g}T^{\mu\nu}{\cal L}_\ep\ph_{\mu\nu}=0
\eeq
where ${\cal L}_\ep$ is the Lie derivative. Either approach
yields
\begin{equation}\label{conlaw}
T^{\mu \nu}_{\ \ \ ; \nu} = {3 \over 2}T^{\alpha \beta}S^{\mu}_{\ \alpha
\beta}
.\end{equation}
This result shows that when the torsion vanishes, we obtain the
conventional result that the covariant
derivative of the energy momentum vanishes.
The equations of motion then follow from this result. With torsion,
(\ref{conlaw}) shows that there will be additional forces due to
torsion, and that geodesic motion should not be expected.

\section{Equations of motion}
\subsection{The general case}
The equations of motion can be found using the method of
Papapetrou.\foot At first, the method will be kept general,
meaning that the actual source will not be specified. After that,
the formulation will be used to find the equation of motion
of a small string in an external field. Some general
comments about this method are: It is not generally covariant.
Volume integrals over the test object are considered at
constant $x_0$. Also, under consideration is the motion of a
small test object in the presence of a large object. The gravitational
field of the test object is essentially ignored, except in the way the
inertial mass is defined. These notions will be clarified as we go.

The starting point is the identity
\beq\label{id}
\tilde T^{\mu\nu}_{\ \ ,\nu}= \sqrt{-g}T^{\mu\nu}_{\ \ \ \ ;\nu}
-\{_{\al \be}^{\; \mu}\} \tilde T^{\al\be}
\eeq
where the tilde implies density according to
$\tilde T^{\mu\nu}=\sqrt{-g}T^{\mu\nu}$.
Now, consider a small volume $d^3x$ that completely enclosed the test
object, and integrate this equation over that volume. In that region, the
energy momentum tensor of the large body is zero, so from here on the
energy momentum tensor is that of the small body---the string.
In other words, we suppose that
\beq
I_m=I_b+I_s
\eeq
where $I_b$ is the material action corresponding to
the energy momentum tensor of the big body, and $I_s$
that of the small body so that
\beq\label{emts}
\de I_s={1\over2}\int d^4x\sqrt{-g}T^{\mu\nu}_s\de \ph_{\mu\nu}
,\eeq
and from here on the subscript $s$ is dropped.

The metric tensor that appears in these formulas is total
gravitational field of both objects, however, assuming that the large
object has much more mass than the small object, the metric tensor
that appears in (\ref{id}) is approximated by that of the large body.
Discarding surface terms, and using the conservation law
(\ref{conlaw}), (\ref{id}) becomes

\beq
{d\over dx^0}\int \tilde T^{\mu0}=
\frac3 2\int\tilde T^{\al\be}S^\mu_{\ \al\be}
-\int\tilde T^{\al\be}\{_{\al\be}^{\; \mu}\}
\eeq
where the volume element has been, and will be, suppressed.
The following definitions will be useful,
with $\ta^{\al\be}\equiv\tilde T^{(\al\be)}$:
\begin{equation}
M^{\mu \nu} = v^{0}\int \tilde T^{\mu \nu}
,\end{equation}

\begin{equation}
M^{\al \mu \nu} = -v^{0}\int \delta x^{\al}\tau^{\mu \nu}
\end{equation}

\begin{equation}\label{m}
m^{\alpha \mu \nu} = v^{0}\int \delta x^{\alpha}\tilde j^{\mu \nu}
\end{equation}

\begin{equation}\label{Jdef}
J^{\mu \nu} =\int (\delta x^{\mu}\tilde T^{\nu o} -\delta x^{\nu}
\tilde T^{\mu 0})
.\end{equation}

We consider $x^\al$ as the coordinate
from the origin to a point on the
small body. Then $y^\al$ is defined according to
$x^\al=y^\al+\de x^\al$, where $\de x^\al<<y^\al$.
To proceed, the Cristoffel symbols and the torsion
tensor are expanded in a Taylor series about the
point $y^\mu$. These quantities may be then taken outside
the integrals, and using the above equations one may show
that
\beq\label{trans0}
{d \over d\tau}\biggl({M^{\mu 0}\over v^{0}}\biggr) +
\{_{\alpha\beta}^{\ \mu}\}M^{\alpha\beta}=
 \{_{\alpha\beta}^{\ \mu}\},_{\eta}
M^{\eta \alpha \beta}
+\frac3 2 M^{\al\be}S^\mu_{\ \al\be}
+3S^{\mu}_{\ \alpha \beta},
_{\eta}m^{\eta \alpha \beta}
.\eeq
In the appendix it is shown that we may take
\beq\label{int0}
\int\tilde j^{\mu\nu}=0
.\eeq
With this

\beq\label{pdef}
p^\mu\equiv{M^{\mu0}\over v^0}=\int\ta^{\mu0}dV
,\eeq
and (\ref{trans0}) becomes

\beq\label{trans1}
{dp^\mu \over  d\tau}+
\{_{\alpha\beta}^{\ \mu}\}M^{\alpha\beta}=
 \{_{\alpha\beta}^{\ \mu}\},_{\eta}
M^{\eta \alpha \beta} +3S^{\mu}_{\ \alpha \beta}, _{\eta}m^{\eta
\alpha \beta} .\eeq
If the torsion vanishes, this equation reduces
to the equation derived by Papapetrou many years ago. The first
term on the right side represents the force on a particle with
structure due the gradient of the field. This gives rise to the
well know result (still no torsion) that only point particles with
no structure follow along geodesics. Any structure to a particle
will give rise to $M^{\al \mu \nu}$, which by (\ref{trans1})
gives rise to non-geodesic motion. Thus we anticipate
that strings, even in Riemannian space,
 will not follow geodesic motion. The second term
on the right side of (\ref{trans1}) is the force on the particle
due to the torsion. It was shown that when the torsion was
constructed from the intrinsic vector, this force was due to the
interaction of the intrinsic spin of the particle and the external
torsion field.

The Papapetrou method also gives rise to the equation for angular
momentum by starting with the identity
\beq
(x^\al\tilde T^{\be\ga})_{,\ga}=\tilde T^{\be\al}
+x^\al\tilde T^{\be\ga}_{\ \ ,\ga}
,\eeq
which gives

\beq
 \int\tilde T^{\be\al}={d\over dx^0} \int(y^\al+\de x^\al)
\tilde T^{\be0}- \int(y^\al+\de x^\al) \tilde T^{\be\ga}_{\ \
,\ga}
.\eeq
Using the same kinds of manipulations as above this
gives
\beq\label{djdt} {d \over d\tau}J^{\alpha
\beta}+{dy^\al\over d\ta}{M^{\beta 0} \over v^{0}} -{dy^\be\over
d\ta}{M^{\alpha 0} \over v^{0}}= 6S^{[\beta}_{\ \mu \nu}m^{\alpha
]\mu \nu} +2\{_{\mu\nu}^{[ \beta}\}M^{\alpha ]\mu \nu}
.\eeq
In the limit that the gravitational and torsional field go to zero
the right side of this equation becomes zero.
In addition, in this limit, (\ref{trans1}),
with (\ref{pdef}), show that $M^{\al0}/v^0$ is the constant
momentum. In this case (\ref{djdt}) yields, upon integration,
\beq\label{jconst}
J^{\al\be}+y^\al p^\be-y^\be p^\al=\mbox {constant}
.\eeq
This shows that,
since the second two terms on the left represent the
orbital angular momentum, $J^{\al\be}$ must represent the total
rotational angular momentum plus intrinsic spin.
With the torsion set equal to zero, these results are identical
to Papapetrou's results.  The main differences we are about to
encounter below are that torsion is not zero, and the energy
momentum tensor is not symmetric, as explained above.
In fact, one may see already that the non-symmetric part of the energy
momentum tensor enters into $J^{\al\be}$ through its definition,
and therefore we see in general, from this result, that
the antisymmetric part of the energy momentum tensor
represents intrinsic spin.

\subsection{Enter the string}

Now we would like to consider that the energy momentum tensor
that describes the particle discussed above is that of a string, so
that, for a simple Nambu-Goto string we assume

\beq\label{stac}
I_s=\mu\int\sqrt{-\ga}d^2\ze+{\mu\et\over2}
\int\sqrt{-\ga}\ps_{\mu\nu}d\si^{\mu\nu}
\eeq
where
\beq
d\si^{\mu\nu}=\ep^{ab}x,_a^\mu x,_b^\nu d^2\ze
\eeq
and $\sqrt{-\ga}\ep^{10}=1$, etc.
The first term is the usual Nambu-Goto action, and is the
conventional way in which the string is introduced
into gravity. In
order to couple something to the torsion potential $\psi_{\mu\nu}$,
an antisymmetric source term must be invoked. A very natural
choice arises with strings, and this is to couple the torsion
potential to the worldsheet area element, as is done above.
This coupling was first used by Kalb and Ramond.\foot
The string coordinates are labeled by $\ze^a$ where $a$
and $b$ range from 0 to 1. Due to the coordinate invariance
of the string action we may choose $\ze^0=x^0$, and call
$\ze^1=\ze$. In curved space, we are not allowed therefore
to choose the conformal gauge, and we won't.
Using this and the definition (\ref{emts})
one obtains

\beq\label{emtst}
T^{\si\nu}={\mu\over\sqrt{-g}}\int
d^2\zeta\sqrt{-\ga}\de(x-x(\ze))x^\si_{,a}x^\nu_{,b}
(\ga^{ab}+\et\ep^{ab})
\eeq
where

\beq\label{gab}
\ga_{ab}=x^\mu_{,a}x^\nu_{,b}g_{\mu\nu}
.\eeq

It is worth emphasizing some differences that arise between these
equations and those that appear in string theory or in the study
of cosmic strings. First, of course, this is a completely
classical presentation. However, we do not impose that variations
of the string action with respect to the coordinate vanish. Thus,
the common `geodesic' condition that is often imposed on the string
coordinates, which in curved space this disallows a static rigid string
(from the Nambu-Goto action), is not enforced.
The equation of motion of the string is derived below
from the Bianchi identity. Also, since we chose $x^0=\ze^0$,
we are not able to arbitrarily choose a gauge for the string
metric $\ga^{ab}$. In this case, we are considering  the equation
of motion of the string in an external field, so $\ga^{ab}$ is
determined from this external field according to (\ref{gab}). In
fact, from (\ref{gab}) we see that $\ga_{00}=(v^0)^{-2}$.  We also
see that $\ga_{01}=g_{01}\pa x^1/\pa\ze$. Assuming that $g_{01}<<
g_{00}, \  g_{11}$, we may ignore the off diagonal components of
the two dimensional metric. Finally, we see that
$\ga_{11}=g_{mn}(dx^m/d\ze)(dx^n/d\ze)$ where $m,n=1-3$. To
calculate this we assume that $g_{mn}$ in the last equation
can be replaced by $\et_{mn}$. To justify this, we should
examine the equation of motion (\ref{trans1}) where this will
be used. If we limit the discussion to weak fields so
that  $g_{\mu\nu}=\et_{\mu\nu}+h_{\mu\nu}$ where
$h_{\mu\nu}<<\et_{\mu\nu}$, then we may be content to carry
only terms linear in $h_{\mu\nu}$ in the equation of motion.
Now, since $T^{\mu\nu}$ (through the quantities $M^{\mu\nu}$
etc.,), which contains $\ga_{ab}$, is multiplied
by $g_{\mu\nu}$ and its derivatives (and products), to this
order it is sufficient to replace
$g_{\mu\nu}$ by $\et_{\mu\nu}$ in (\ref{emtst}). With this,
in the appendix it is shown that
$\ga_{11} =-1$, so that
 the energy momentum tensor of the string becomes

\beq\label{emtmink}
\tilde T^{\al\be}={\mu\over v^0}
\int d^2\ze\de(x-x(\ze))[(v^\al v^\be-x'^\al x'^\be)
+\et(v^\be x'^\al-v^\al x'^\be)]
\eeq
where $v^\al=dy^\al/d\ta$ is the four velocity of the center of mass
and $x'^\al=dx^\al/d\ze$.

With this, we can put the equation of motion (\ref{trans1})
can be put into a more recognizable, or useful, form.
For the present we will content ourselves to obtain
the equation of motion in the lowest order. This means that the terms
of the right hand side of (\ref{trans1}) will be neglected
for now. A more detailed examination of these terms will be
reserved for future work.
With (\ref{emtmink}) one obtains

\beq
M^{\al\be}=\mu\int d\ze(v^\al v^\be-x'^\al x'^\be)
.\eeq
With this, (\ref{trans1}) becomes

\beq\label{eqm2}
{d\over d\ta}(mv^\si)+\{^{\ \si}_{\al\be}\}M^{\al\be}=0
.\eeq
where

\beq\label{mass}
m\equiv {M^{00}\over (v^0)^2}=\mu\int d\ze
.\eeq
Now, (\ref{eqm2}) becomes

\beq\label{eqm3}
{dm\over d\ta}v^\si +m {dv^\si\over d\ta}
+\{^{\ \si}_{\al\be}\}\left( mv^\al v^\be-
\mu \int d\zeta x'^\al x'^\be \right)=0
.\eeq
Also defining
\beq
{Dv^\si\over d\ta}={dv^\si\over d\ta} +
\{_{\mu \nu}^{\; \si}\}v^\mu v^\nu
,\eeq
we can use the identity $v_\si Dv^\si/d\ta=0$, so that (\ref{eqm3})
gives

\beq
{dm\over d\ta}=\mu v_\si\{^{\ \si}_{\al\be}\}
\int d\ze x'^\al x'^\be
\eeq
which puts us in the momentarily awkward position
that, even neglecting the terms on the right
side of (\ref{trans1}), we do not have geodesic
motion and the inertial mass is not conserved.
However, consider
\beqa
\int_0^L d\ze x'^\al x'^\be=
\int dx^\al x'^\be
=-\int x^\al d(x'^\be)=\\ \nonumber
-\int(y^\al+\de x^\al) d(x'^\be)\approx-y^\al\int
d(x'^\be)=0
\eeqa
where integration by parts was used and it is assumed that
{\em the string is a closed loop} so that the end
point terms contribute nothing,
and the last step follows for a closed loop that has
no kinks. This shows, finally, the inertial mass is conserved
and that (\ref{eqm3}) reduces to

\beq
 {dv^\si\over d\ta}
+\{^{\ \si}_{\al\be}\}v^\al v^\be=0
.\eeq

This result is received as good news for several reasons. First,
of course, the inertial mass is conserved. Second, in this lowest
order approximation, the string moves along a geodesic.
We also see that the structure of the string will cause
its actual motion to deviate from the geodesic, and these
effects can be calculated from (\ref{trans1}). These results
also show that a sensible equation of motion results from the
Bianchi identity (or from the conservation laws) as it should.
Thus, there is no need to make the additional ``geodesic''
postulate, that variations of the matter action
with respect to the coordinates vanish. In fact, they probably
do not.

To understand better the physical significance
of this energy momentum tensor, and in particular to
see that this implies that strings have intrinsic spin,
we  start from the  spin vector, defined by

\beq\label{spin0}
S_\ga={v^\si\over2} J^{\al\be}\ep_{\si\\al\be\ga}
,\eeq
which for low velocities becomes
 \beq\nonumber
S_\ga\ra\frac1 2J^{\al\be}\ep_{0\al\be\ga}
,\eeq and make the same
approximations for $T^{\al\be}$ as we did above. When
(\ref{emtmink}) is used in (\ref{Jdef}), which is used in
(\ref{spin0}), two kinds of terms arise, those that depend on
velocity (3-velocity) and those that do not. The velocity
dependent terms represent a conventional $\bm{r}\times\bm{p}$
angular momentum of string, about its center of mass, due to its
oscillations or rotations. If we restrict our attention to the rigid
loop, these terms vanish. Even if the string is not rigid, but
is small, i.e., elementary
particle size, this term is negligible. Thus, we restrict
our attention to the terms that do not depend on
velocity, and obtain, for the  circular loop,

\beq\label{spinloop}
S_\ga=2\int d\ze\ep_{0\al\be\ga}\de x^\al{dx^\be\over d\ze}
,\eeq
or

\beq\label{spinloop3}
{\bm S}=\mu\et\int\bm{r}\times\bm{r}'d\ze\ \ \Rightarrow
S=2\mu\et\times\mbox{Area}
\eeq where ${\bm r'}=d{\bm r}/d\ze$, and
${\bm r}$ is a vector from the center the center of mass to a
point on the string.

This result shows that strings, due to the fact that they have structure,
give rise to intrinsic spin when they are coupled to the torsion
potential.
(One may note that the conventional method of discussing
a spinning string in string theory is described by the introduction
the Grassmannian $\ps$, leading to the Raymond model
or the Neveu-Schwarz.\foot
Here, we will have intrinsic spin from the Nambu-Goto action alone,
due only to its structure.)
 Now we may turn our attention to the torsion
field they generate.

\section{Torsion from a string}

The adoption of the string action (\ref{stac}) has two significant
consequences. One, as shown above, ties intrinsic spin to the string.
The other consequence, and in fact the original motivation in
adopting (\ref{stac}), is that it acts a source of torsion.
In this section, we shall consider the Minkowski
limit, and solve the torsional field equations in this limit.
Of course, since torsion enters in the gravitational field
equations, strictly speaking, torsion cannot exist without
a concomitant gravitational field. However, as has been shown, when
torsion arises from intrinsic spin the torsion field is very small
and Minkowski spacetime is a very good approximation.

It is helpful to use the dual to torsion

\begin{equation}\label{bmu}
b_{\mu} = \epsilon_{\mu \alpha \beta \gamma}S^{\alpha \beta \gamma}
\end{equation}
and $\bm{b}=(b_n)$.
With this in mind (\ref{tfe}) becomes

\beq \Box\ps^{\mu\nu}=-3k j^{\mu\nu}
.\eeq
Now we consider the static case, so that this reduces to
\beq
\label{lap} \na^2\bm{A}=\bm{N}
\eeq where
\beq
A_n\equiv2\ps_{0n},\ \ \ \ \ \ \bm{A}=(A_n)
\eeq and
\beq
N^n\equiv6kj^{0n},\ \ \ \ \ \ \bm{N}=(N^n)
\eeq and a  `Lorentz'
gauge is chosen so that \beq \ps^{\si\mu}_{\ \ ,\si}=0
,\eeq
which is allowed due to the gauge invariance
$\ps_{\mu\nu}\ra\ps_{\mu\nu}+\xi_{[\mu,\nu]}$.
Ordinary Green's function techniques may now be used to give the
solution to ({\ref{lap}) as
\beq
 \bm{A}(x)={1\over4\pi}\int
G(\bm{x}-\bm{y})\bm{N}(\bm{y})d^3y
\eeq where
\beq
G={1\over|\bm{x}-\bm{y}|}=\frac1 x+{\bm{x}\cdot\bm{y}\over
x^3}+...
\eeq

\no To lowest order (using only the $1/x$ term in the
expansion) gives
\beq A_n={3k\over2\pi x}\int  j^{0n}d^3y
\eeq
where
\beq j^{0n}={\et\mu\over2}\int
d^2\ze\de^4(x-x(\ze))x_{,a}^0x_{,b}^n\ep^{ab}
\eeq so

\beq \int
{2\over\et\mu}j^{0n} d^3y=\int_0^Ldx^n=0
\eeq
for closed strings. Now, using the
next term in the expansion we have
\beq A_n={\bm {x}\over4\pi
x^3}\cdot\int\bm{y}N^nd^3y
.\eeq

\no In these formulas $\bm{y}$ is
the body (string) centered coordinate, and represents the spatial
part of $\de x^\mu$ used above. Using (\ref{spinloop3}) the
solution be comes \beq
\bm{A}={3k\over8\pi}{\bm{S}\times\bm{x}\over x^3}
.\eeq
With this, the torsional field equations yield,
letting ${\bm r}$ replace  ${\bm x}$,
and considering the case that the spin points in the $z$ direction,
\beq\label{torsol} {\bm
b} = {3k\over8\pi}{S\over r^{3}} \left(2 \cos(\theta) \bm {\hat r} +
\sin(\theta)\bm{\hat \theta} \right)
.\eeq}
The physical significance
of this result is that the intrinsic spin, which results from
the structure of the string, gives rise to a dipole field---the torsion
field.

\section{Conclusions}
A string not only acts as the source of a gravitational field,
it also becomes the source of torsion. In fact, when the string
worldsheet area is coupled to the torsion potential, the physical
property of the string that gives to torsion is intrinsic spin.
The work presented here focused on closed strings only. The
intrinsic spin of the string does not arise from motions of the
string, but is due to the structure and spatial extent of the  string.
The Bianchi identity may be used to find the equation of motion
of the string, and we found that the motion is geodesic
only in lowest order.

\section{Appendix}

Above, the volume integral of the torsion source
tensor was taken to vanish. This was done in (\ref{int0}).
This can be shown to be true for two situations. One corresponds
to the case that the string is a static rigid circle.
It is noted that this assumption cannot hold for large cosmic
strings under the influence of their own gravity.
This result comes from the equation of motion of string, which is obtained
by taking variations with respect to $x^\mu$. However, this `geodesic'
postulate is not adopted here, the equation of motion of the string
is obtained from the Bianchi identities, so this restriction does
not hold. Allowing oscillations of the string nevertheless,
we may then show that the time average of the integral will vanish.
Assuming the period is very small, this is just as good as the rigid
string assumption.

First, without any assumptions, one may see that,
putting back the 3-volume element
\beq
\int\tilde j^{\si0}d^3x=0
.\eeq
From the definition $j^{\mu\nu}\equiv (1/2)T^{[\mu\nu]}$ and
(\ref{emtst}) we have

\beq
j^{\si\nu}={\mu\et\over2\sqrt{-g}}\int
d^2\zeta\sqrt{-\ga}\de(x-x(\ze))x^\si_{,a}x^\nu_{,b}\ep^{ab}
.\eeq
Integrating this over a volume one has,

\beq
\int\tilde j^{\si0}d^3x={\et\mu\over2}\int_0^Ld\ze{dx^\si\over d\ze}=0
\eeq
where again this holds for closed strings.

Now it is shown that $\int\tilde j^{mn}d^3x=0$ for a rigid loop, or
on average. For a rigid loop this becomes

\beq
\int\tilde j^{mn}d^3x=
-{\et\mu\over2}\int(y^m+\de x^m)d\left({dx^n\over dx^0}\right)
\eeq
after integration by parts. The first term vanishes since $y^m$
goes to the center of mass and if $d(\de x^n)/dx^0=0$,
as it would for a rigid string, then the second term vanishes too,
so that we have shown

\beq
\int \tilde j^{\mu\nu}=0
.\eeq

Alternatively, we may allow the string to undergo periodic
oscillations in which case it is to be shown that

\beq
<\int j^{\mu\nu}d^3x>\equiv\frac1 T\int dx^0\int j^{\mu\nu}d^3x=0
.\eeq
In this case we have
\beq
<\int j^{\mu\nu}d^3x>=
{\et\mu\over T}\int dx^0d\ze\left({dx^\nu\over dx^0}{dx^\mu\over d\ze}
-{dx^\mu\over dx^0}{dx^\nu\over d\ze}\right)=0
.\eeq
Finally, one may see that  the average over one period
of the symmetric part of the energy momentum tensor does not
vanish, and essentially gives back the energy momentum tensor,
as we would expect.

Now we show the same kind of thing for the spatial part of the
worldsheet metric. First, if we assume that the string is a rigid
circular loop and, defining the $\ph$ as an angular measure as
$\de x^\mu$ traverses the loop, we get $\ga_{11}
=-(\cos^2\ph+\sin^2\ph)=-1$. On the other hand, if again we assume
that there are periodic oscillations it is shown that $<\ga_{11}>$
is approximately equal to -1. To see this, note that
 \beq
<\ga^{11}>=\frac1 T\int dx^0{dx^m\over d\ze}{dx^n\over d\ze}g_{mn}
.\eeq
Now use $x^m=y^m+\de x^m$ and assume further that $\de x^m$
may be written as $\de x^m=a^m+\ep^m(t)$ where $a^m$ is
time independent and $\ep^m$ is periodic in $T$. Then,
\beq
<\ga_{11}>\approx \frac1 T\int dx^0{d\over d\ze}(a^m+\ep^m)
{d\over d\ze}(a^n+\ep^n)\et_{mn}
.\eeq
Now, retaining terms to order $\ep^m$ and using the periodicity
of $\ep^m$ we have
\beq
<\ga_{11}>={d a^m\over d\ze}{d a^n\over d\ze}\et_{mn}
+{2\over T}{d a^m\over d\ze}{d \over d\ze}\int_0^Tdx^0\ep^n\et_{mn}
.\eeq
The second terms is zero due to the periodicity and the first
terms replicates the result for the rigid loop, so we have
\beq
<\ga^{11}>=-1
.\eeq

\end{document}